\documentstyle[11pt]{article}

\makeatletter
\def\dif{{\rm d}}
\def\deriv{\@ifnextchar[{\@deriv}{\@deriv[]}}
   \def\@deriv[#1]#2#3{\mathchoice%
{{\dif^{#1}#2\over\dif{#3}^{#1}}}{{\dif^{#1}#2/\dif{#3}^{#1}}}%
{{\dif^{#1}#2\over\dif{#3}^{#1}}}{{\dif^{#1}#2/\dif{#3}^{#1}}}}
\def\derpar#1#2{\mathchoice%
{{\partial#1\over\partial#2}}{{\partial#1/\partial#2}}%
{{\partial#1\over\partial#2}}{{\partial#1/\partial#2}}}
\def\dderpar#1#2#3{\mathchoice%
{{\partial^2 #1\over\partial #2\,\partial #3}}%
{{\partial^2 #1/\partial #2\,\partial #3}}%
{{\partial^2 #1\over\partial #2\,\partial #3}}%
{{\partial^2 #1/\partial #2\,\partial #3}}}
\def\secteqno{\@addtoreset{equation}{section}%
\def\theequation{\thesection.\arabic{equation}}}
\newcounter{subequation}
\def\thesubequation{\alph{subequation}}
\def\sneqnarray{\stepcounter{equation}\let\@currentlabel=\theequation
\setcounter{subequation}{1}
\def\@eqnnum{{\rm (\theequation.\thesubequation)}}
\global\@eqcnt\z@\tabskip\@centering\let\\=\@eqncr\let\@@eqncr=\@@sneqncr
$$\halign to \displaywidth\bgroup\@eqnsel\hskip\@centering
 $\displaystyle\tabskip\z@{##}$&\global\@eqcnt\@ne
 \hskip 2\arraycolsep \hfil${##}$\hfil
 &\global\@eqcnt\tw@ \hskip 2\arraycolsep $\displaystyle\tabskip\z@{##}$\hfil
  \tabskip\@centering&\llap{##}\tabskip\z@\cr}
\def\endsneqnarray{\@@sneqncr\egroup $$\global\@ignoretrue}
\def\@@sneqncr{\let\@tempa\relax
   \ifcase\@eqcnt \def\@tempa{& & &}\or \def\@tempa{& &}
   \else \def\@tempa{&}\fi
     \@tempa \if@eqnsw\@eqnnum\stepcounter{subequation}\fi
     \global\@eqnswtrue\global\@eqcnt\z@\cr}
\def\ben{\begin{enumerate}}
\def\een{\end{enumerate}}
\def\beq{\begin{equation}}
\def\eeq{\end{equation}}
\def\bea{\begin{eqnarray}}
\def\eea{\end{eqnarray}}
\def\beann{\begin{eqnarray*}}
\def\eeann{\end{eqnarray*}}
\def\beasn{\begin{sneqnarray}}
\def\eeasn{\end{sneqnarray}}

\newtheorem{teor}{Theorem}

\def\tabaddress#1{{\small\it\begin{tabular}[t]{c}#1\\[1.2ex]\end{tabular}}}

\def\UBECM{Departament d'Estructura i Constituents de la Mat\`eria\\
   Universitat de Barcelona,\\
and Institut de F\'\i sica d'Altes Energies,\\
   Av.~Diagonal 647\\
   08028 Barcelona\\
   Catalonia, Spain}
\def\UPCMAT{Departament de Matem\`atica Aplicada i Telem\`atica\\
   Universitat Polit\`ecnica de Catalunya\\
   Campus Nord UPC, edifici C3\\
   C. Jordi Girona, 1\\
   08034 Barcelona\\
   Catalonia, Spain}

\let\ds=\displaystyle

\def\Real{{\bf R}}
\def\Tan{{\rm T}}
\def\FL{\hbox{$\cal F$}\!L}
\def\deltaH{\delta^{^{\rm H}}\!}
\def\deltaL{\delta^{^{\rm L}}\!}
\def\GH{G^{\!^{\rm H}}}
\def\GL{G^{\!^{\rm L}}}

\def\phc{\hbox{\sf phc}}
\def\plc{\hbox{\sf plc}}

\let\oldGamma=\Gamma
\def\Gamma{{\bf\oldGamma}}
\def\VL{{\bf V}^{\!^{\rm L}}}
\def\VH{{\bf V}^{\!^{\rm H}}}
\def\XL{{\bf X}^{\!^{\rm L}}}
\def\XH{{\bf X}^{\!^{\rm H}}}
\def\K{{\bf K}}

\def\V{{\bf V}}
\def\X{{\bf X}}
\def\Z{{\bf Z}}

\def\bfp{{\bf p}}
\def\bfx{{\bf x}}

\def\eps{\varepsilon}
\def\dddot#1{\hbox{$\mathop{#1}\limits^{\ldots}$}}

\def\relstack#1#2{\mathrel{\mathop{#1}\limits_{#2}}}
\def\feble#1{\relstack{\approx}{#1}}

\title{Canonical Noether symmetries and commutativity properties\\
for gauge systems%
\thanks{published: 
{\sl J.~Math.\ Phys.~\bf 41} (2000) 7333--7351}
}

\author{Xavier Gr\`acia$^a$ and Josep M. Pons$^b$
\\[2mm]
\tabaddress{$^a$\UPCMAT}
\\
\tabaddress{$^b$\UBECM}
\\[2mm]
\small emails: 
{\tt xgracia@mat.upc.es}, 
{\tt pons@ecm.ub.es}
}

\date{17 May 2000 / updated April 2001}

\begin{document}

\maketitle

\begin{abstract}
\noindent
For a dynamical system defined by a singular Lagrangian,
canonical Noether symmetries are characterized in terms of their
commutation relations with the evolution operators
of Lagrangian and Hamiltonian formalisms. 
Separate characterizations are given in phase space, in velocity space,
and through an evolution operator that links both spaces.

\bigskip
\parindent 0pt
\it


PACS\,1999: 45.20.Jj, 02.30.Wd, 11.30.-j
\qquad
MSC\,2000: 70H33, 70H45, 34C14

\end{abstract}

\clearpage
\section{Introduction}

Most physical theories implement the dynamics as a result of 
the application of a variational principle, that is, by means of a 
Lagrangian. 
Among the dynamical symmetries of these theories, that is, transformations 
that map solutions of the equations of motion into solutions,
we can single out the Noether symmetries, 
that is, the continuous transformations that leave the action invariant 
---except for boundary terms.
In addition, 
if we aim to move the description of the dynamics  
from the tangent bundle (velocity space) $\Tan Q$ 
of its configuration space~$Q$ 
to the cotangent bundle (phase space) $\Tan^*Q$, 
other distinctions can be raised, 
as to whether the symmetry transformation in velocity space
is projectable to phase space and, 
in the affirmative case, 
whether the transformation in phase space is canonical.
We will consider time-independent Lagrangians,
as it is the usual case in physical theories,
but we will allow to deal with time-dependent functions to cover 
also gauge symmetries (symmetries depending upon arbitrary functions of time, or space-time variables in field theory);
then we will use $\Real \times \Tan Q$ and $\Real \times \Tan^*Q$
instead of $\Tan Q$ and $\Tan^*Q$.

The infinitesimal symmetries of an ordinary dynamical system are 
characterized by a property of commutativity: 
essentially, that the time evolution operator commutes with  
the operator that generates the symmetry.
Let us state with more detail this result,
which is standard for theories with no gauge freedom, using 
differential-geometric language. 
Let $\X$ be the vector field that governs the dynamics 
(the time evolution) 
of some system on a given manifold ${\cal M}$
(${\cal M}$ can be, for instance, 
$\Real \times \Tan Q$ or $\Real \times \Tan^*Q$
for some configuration manifold~$Q$; 
$\Real$ parametrizes the independent variable ---the time). 
For an open interval $I \subset \Real$, a path 
$
\gamma \colon I \to {\cal M}
$
is a solution to the dynamics if 
$\dot \gamma = \X \circ \gamma$. 
Let a vector field $\V$ be a candidate for a symmetry of the dynamics 
defined by~$X$.
Then the flow of~$\V$ (a local one-parameter group of diffeomorphisms) 
transforms solutions into solutions 
if and only if $\X$ is $\V$-invariant,
that is to say,
\beq
{\cal L}_\V \X = [ \V, \X ]= 0 ,
\label{purecomm}
\eeq
where ${\cal L}_\V$ stands for the Lie derivative. 
This is an immediate consequence of the well-known fact that
$[\V,\X]= 0$ iff their flows commute
\cite{Olv93-diffeq,AM-mec,Arn-mec}.

%

Our aim in this paper is to obtain some generalized versions of this result.
More precisely,
our purpose is to study how the canonical Noether transformations
implement this commutativity requirement 
in the general case of gauge theories 
(those derived from singular Lagrangians).
Instead of providing with new procedures to determine symmetries, 
we give alternative ways to characterize them, 
associated with a specific property of commutativity. 
Recall that the variation of the Lagrangian under a Noether symmetry 
is a total derivative;
this statement is far from expressing any kind of commutativity. 
We will discover however 
that one can characterize canonical Noether symmetries through 
commutativity properties; in this way, we give a new perspective, with a 
geometrical flavor, 
to identify the Noether symmetries of a dynamical system. 
This approach can be applied in particular to gauge theories,
where it can be used as a direct test as to whether a given transformation
is a Noether symmetry.

Since many dynamical systems
---and, among them, those describing the fundamental inter\-actions---
have room for gauge freedom, 
we will assume in our framework that the Lagrangian may be singular. 
To be more concrete: we will consider theories described by 
time-independent first-order Lagrangians whose
Hessian matrix with respect to the velocities may be singular. 
In this case the conversion from tangent space language to phase
space language has some peculiarities: 
there are constraints in the formalism, 
the dynamics has some degree of arbitrariness, etc.
This is nothing but the framework first studied by Dirac to deal with 
gauge theories or, more generally, constrained systems
\cite{Dir50,Dir64,BGPR86-equiv,Car90-theory,MMS-constraints,HRT76}.
The regular case is recovered when no Hamiltonian constraints occur.
 
Throughout the paper we will only consider continuous symmetries.
Among them,
how can we distinguish the Noether symmetries?
The distinction comes in part from the following fact: 
a Noether symmetry has an associated conserved quantity, 
and this conserved quantity contains all the information to reconstruct 
the symmetry
\cite{Olv93-diffeq}.
This fact characterizes a Noether symmetry for regular Lagrangians
(those with regular Hessian matrix), 
but not in the general case of gauge theories that we are also addressing: 
there are symmetries with conserved quantities that are not Noether.

Let us distinguish clearly the singular case from the regular one.
In the regular case we know that:
\vadjust{\kern-2mm}
\begin{itemize}
\itemsep 0pt
\parskip 0pt plus 1pt
\item[(a)]
There is a one-to-one correspondence between Noether 
symmetries and conserved quantities.
\item[(b)]
When formulated in phase space, the conserved quantities become the
generators, through the Poisson bracket, of the Noether symmetries. 
Therefore, Noether symmetries are canonical transformations.
\end{itemize}
Instead, in the case including gauge theories, 
we can list a very different set of assertions:
\vadjust{\kern-2mm}
\begin{itemize}
\itemsep 0pt
\parskip 0pt plus 1pt
\item[(a)]
There can be conserved quantities in phase space that do not generate
symmetries at all.
\item[(b)]
There can be conserved quantities in phase space that generate
symmetries that are not Noether. 
\item[(c)]
There can be nontrivial Noether symmetries whose conserved quantity
in velocity space is identically vanishing.
\item[(d)]
There can be Noether transformations in tangent space that are not
projectable to phase space 
(but the conserved quantity is always projectable).
\item[(e)]
It remains true that, regardless as to whether the Noether symmetry 
is projectable or not to phase space, it can be always reconstructed 
through the Poisson bracket by
using the conserved quantity in phase space. 
In other words, the conserved
quantity still encodes all the information to reconstruct the symmetry.
\item[(f)]
When the Noether symmetry is projectable to phase space,
it is also true that such symmetry is always a canonical transformation 
that is generated by a conserved quantity.
We call such a symmetry a canonical Noether transformation.
\end{itemize}

Let us briefly comment on these assertions.

To prove (a) it suffices to realize that any second class constraint is
a conserved quantity that does not generate a symmetry: it takes the
motions out of the constraint surface.

Statement (b) is a consequence of the fact that 
the conserved quantities $\GH$ that generate canonical Noether transformations 
satisfy stricter conditions 
($\K \cdot \GH = 0$, 
see equation (\ref{KG}) in section~3)
than the ones required to generate dynamical symmetry 
transformations in phase space
($\K \cdot \GH = \,\hbox{quadratic constraints}$, 
see ref.~%
\cite{GP88-ggen});
this is illustrated at the end of the second example in section~4.

The occurrence of (c) is studied in 
\cite{GP94-vanish},
and it happens when the number of independent primary Lagrangian constraints 
is less than the number of independent primary Hamiltonian constraints;
the simplest example is given by the free relativistic particle,
that does not have Lagrangian constraints.

An example of statement (d) is provided, 
in any time-independent gauge theory, 
by the Noether symmetry associated with time translations:
the variation $\delta q = \dot q$ is not projectable to phase space, 
whereas its conserved quantity, the energy, 
projects to the Hamiltonian function.
The projectability of the conserved quantity 
associated with any Noether transformation was noticed in
\cite{Kam82-constr}.
On the other hand, special situations may often arise
when studying the projectability of the gauge transformations,
as for example
the non existence of Hamiltonian gauge generators
of a certain model possessing Lagrangian gauge transformations
\cite{GP92-hamnoeth},
and the loss of covariance of the Hamiltonian gauge transformations
for a particle model admitting a Lorentz covariant Hamiltonian formulation
\cite{GR93-conformal}.

Statement (e) is explained in
\cite{GP92-hamnoeth} and
\cite{GP99-noeth},
where several examples can be found.
Finally, assertion (f) is proven in 
\cite{BGGP89-noether}.

\medskip

From these considerations, we see that it is important to characterize 
the conserved quantities,
because they already encode the transformation. 
This is the usual procedure when one considers Noether symmetries.   
In this paper we propose a shift of emphasis: 
instead of focusing on the conserved quantities, 
we will be interested in properties of the transformations themselves. 
We will show the relevance of commutation properties 
in order to characterize Noether symmetries. 
In this sense, from a theoretical viewpoint 
we will enlarge the list of properties above; 
from a practical viewpoint 
we will provide with new instruments to check whether a given
transformation is a Noether symmetry.

\medskip

We organize the paper as follows.
The basic notations and some preliminary results are set up in section~2.
Section 3 is mainly devoted to the study of Noether transformations
that are projectable to phase space;
these transformations are given different characterizations in terms
of commutation relations involving the evolution operators 
of the Hamiltonian and the Lagrangian formalisms.
Section 4 contains some examples illustrating these results,
and section 5 is devoted to conclusions.

\section{Notation and preliminary results}

We consider a configuration space~$Q$,
with velocity space the tangent bundle $\Tan Q$,
and a (time-independent, first-order) Lagrangian function
$L(q,\dot q)$ defined on it.
The fiber derivative of~$L$ defines the Legendre's transformation,
which is a map
from velocity space to phase space,
$\FL \colon \Tan Q \to \Tan^*Q$,
locally defined by
$$
\FL(q,\dot q) = (q,\widehat p) ,
$$
where we have introduced the momenta
$\widehat p = \derpar{L}{\dot q}$
---we will suppress most indices.

Given a function $h(q,p)$ in phase space,
its pull-back
(through the Legendre's transformation $\FL$)
is the function $\FL^*(h)$ in velocity space
obtained by substituting the momenta by their
Lagrangian expression:
$
\FL^*(h) (q,\dot q) = h \left(q,\widehat p \right)
$.
A function $f(q,\dot q)$ in velocity space is called
$\FL$-projectable ---or, simply, projectable---
if it is the pull-back of a certain function
$h(q,p)$.

\smallskip
We shall always assume that the Legendre's transformation 
$\FL$ has constant rank;
this amounts to say that the fibre Hessian of~$L$, 
which is locally described by 
the Hessian matrix with respect to the velocities
$$
W = \dderpar{L}{\dot q}{\dot q} ,
$$
has constant rank.
Notice that gauge symmetries can only exist when this rank
is not maximal;
this is the case we are interested in.

Let $\gamma_{\mu}$ 
($\mu=1,\ldots ,p_0$)
be a basis of the null vectors of~$W$;
then  the necessary and sufficient condition
for a function $f(q,\dot q)$ in $\Tan Q$
to be (locally) projectable to $\Tan^*Q$ is
\beq
\Gamma_{\mu} \cdot f = 0
\label{projectable}
\eeq
for each~$\mu$,
where the vector fields
$\ds \Gamma_{\mu} := \gamma_{\mu} \derpar{}{\dot q}$
indeed span a basis of the kernel of the tangent map $\Tan(\FL)$.

Under the same assumption about the constant rank,
the image $P_0$ of the Legendre's map
can be locally taken as the submanifold of phase space
described by the vanishing of $p_0$
primary Hamiltonian constraints~$\phi_{\mu}$,
linearly independent at each point of~$P_0$. 
So they satisfy $\FL^*(\phi_\mu) = 0$ by definition.
Then the basis $\gamma_{\mu}$ can be taken as
\cite{BGPR86-equiv}
\beq
\gamma_{\mu} := \FL^* \left( \derpar{\phi_{\mu}}{p} \right) .
\label{gamma}
\eeq

\medskip
Though our Lagrangian is time-independent,
we will need to consider time-dependent functions.
The adjunction of the $t$-variable where needed will not 
cause any problem.
The time-derivative operator acting on a function $f(t,q,\dot q)$ is
$$
\deriv{}{t} =
\derpar{}{t} + \dot q \,\derpar{}{q} + \ddot q \,\derpar{}{\dot q} ,
$$
with the acceleration $\ddot q$ as an independent variable
(this involves the tangent bundle of second order, $\Tan^2Q$).
Then the Euler-Lagrange equations can be written
$$
[L]_{(q,\dot{q},\ddot{q})} = 0 ,
$$
where we have defined
\beq
[L] :=
\derpar{L}{q} - \deriv{\widehat p}{t} =
\alpha - \ddot q W ,
\label{L}
\eeq
with
$\displaystyle  \alpha = \derpar{L}{q} - \dot q \dderpar{L}{q}{\dot q}$.
The primary Lagrangian constraints arise from it,
\beq
\chi_\mu := \alpha \, \gamma_\mu = [L] \, \gamma_\mu ,
\label{primlag}
\eeq
though they are not necessarily independent;
their vanishing defines a subset $V_1 \subset \Tan Q$.

As a matter of notation,
it is usual to write
$f \feble{M} 0$ to mean that $f(x)=0$ for all $x \in M$
(Dirac's weak equality);
for instance
$\phi_\mu \feble{P_0} 0$ and
$\chi_\mu \feble{V_1} 0$.

\medskip
In a gauge theory the dynamics either in Lagrangian or Hamiltonian
formalisms has a certain degree of arbitrariness.
One can introduce
a useful differential operator $\K$  
connecting the Lagrangian and Hamiltonian formalisms,
that has no ambiguity at all, and that still represents the dynamics
\cite{BGPR86-equiv}.
It can be defined as a vector field along 
the Legendre's transformation $\FL$
\cite{GP89-evol},
and, as a differential operator,
it gives the time evolution of a function 
$h$ in $\Real \times \Tan^*Q$ 
as a function $\K \cdot h$ in $\Real \times \Tan Q$ by
\beq  
\K \cdot h := 
\FL^*\left(\derpar ht\right) +
\FL^*\left(\derpar hq\right) \dot q +
\FL^*\left(\derpar hp\right) \derpar Lq .
\label{K}
\eeq
The operator $\K$ is directly determined by the Lagrangian by just taking
partial derivatives. 
Instead, the determination of the dynamics either
in tangent space or in phase space requires more involved computations. 
In this sense, $\K$ is the simplest among the evolution operators,
and this will turn out to be advantageous in order to characterize the Noether symmetry transformations 
by way of commutativity properties.
The operator $\K$ is especially valuable in the study of singular Lagrangians.
For instance, all the Lagrangian constraints are obtained
by applying it to the Hamiltonian constraints
\cite{Pon88-newrel},
and the Lagrangian and Hamiltonian dynamics can be described geometrically
by using this operator
\cite{GP89-evol}.
The operator $\K$ will be instrumental in obtaining some of the results
of the next section.

It will prove very convenient to present two other 
equivalent expressions for the operator~$\K$,
to be used in the next section.
The first one is
\beq
\K \cdot h = 
\deriv{}{t}\FL^*(h) + [L]\, \FL^*\left( \derpar hp \right) ,
\label{K-EL} 
\eeq
whose proof is direct by using the chain rule
\cite{GP92-hamnoeth}.
A direct consequence of this equation and definition (\ref{primlag}) is 
another expression for the primary Lagrangian constraints:
\beq
\chi_\mu = \K \cdot \phi_{\mu} .
\label{primlag'}
\eeq

The second expression relates $\K$ with the Hamiltonian evolution
\cite{BGPR86-equiv}:
\beq
\K \cdot h =
\FL^*\left( \derpar{h}{t} \right) +
\FL^*\{h,H\} +
\sum_\mu \FL^*\{ h,\phi_\mu \} \,v^\mu .
\label{K-H}
\eeq
Here $H$ is any Hamiltonian function
(its pull-back to $\Tan Q$ is the Lagrangian energy;
it is defined up to primary Hamiltonian constraints).
And the $v^\mu(q,\dot q)$
are functions uniquely determined by this equality
when one takes $h = q^i$;
these functions are not projectable, and indeed
\beq
\Gamma_\nu \cdot v^\mu = \delta^\mu_{\,\nu} .
\label{Gamlam}
\eeq
A consequence of (\ref{K-H}) is a test of projectability
for the function $\K \cdot h$:
\beq
\Gamma_\mu \cdot (\K \cdot h)  = \FL^* \{ h,\phi_\mu \} ,
\label{Gamma-K}
\eeq
so $\K \cdot h$ is projectable iff $h$ is a first-class function
with respect to the primary Hamiltonian constraint submanifold~$P_0$.

The Lagrangian time-evolution differential operator
can be expressed 
\cite{BGPR86-equiv}
as
\beq 
\XL = 
\XL_0 + \eta^{\mu} \Gamma_{\mu} ,
\label{lagrevol}
\eeq
where the $\eta^{\mu}$ are in principle arbitrary functions 
that express the gauge freedom of the time-evolution operator
and  $\XL_0$ is a vector field in velocity space
\beq
\XL_0 =
\derpar{}{t}
+ \dot q^{i} {\partial\over\partial q^{i}}
+ a^{i}(q,\dot q) {\partial\over\partial \dot q^{i}} .
\label{X0}
\eeq
The accelerations $a^i$ in $\XL_0$ may be determined by the formalism,
with some arbitrariness owing to the gauge freedom,
and we do not need here their explicit expression, 
which is given in
\cite{BGPR86-equiv}.
The nature of this operator has been recently discussed in
\cite{Gra-fibder,GP-newstruc}.
In view of application we only need to know its relationship 
with the operator $\K$
\cite{Pon88-newrel}:
\beq
\K \cdot h = 
\XL_0 \cdot \FL^*(h) + 
\chi_\mu \, \derpar{v^\mu}{\dot q} \, \FL^*\left( \derpar{h}{p} \right) .
\label{XandK}
\eeq

\section{Canonical Noether transformations for gauge theories}

Now we are ready to study the symmetries in Lagrangian and Hamiltonian
formalisms as commutation relations between these symmetries and the dynamics.
The case of gauge theories will lead to modified versions of 
equation (\ref{purecomm})
that account for the existence of constraints
and the ambiguity of the dynamics due to gauge freedom.

Let us consider an infinitesimal Noether transformation
$\deltaL q(t,q,\dot q)$ in configuration space,
that is to say, 
the variation of $L$ is a total time-derivative.
Then a conserved quantity $\GL$ arises:
\begin{equation} 
[L]_i \,\deltaL q^i + \deriv{\GL}{t} = 0 .
\label{noether}
\end{equation}
As we have recalled in the introduction,
the conserved quantity is always projectable 
\cite{Kam82-constr} 
to a function $\GH(t,q,p)$ in phase space,
$\GL = \FL^*(\GH)$.
This is proved by extracting the coefficient of the acceleration $\ddot q$
from equation (\ref{noether})
and then saturating the result with the null vectors $\gamma_\mu$
of the Hessian matrix~$W$,
thus obtaining $\Gamma_\mu \cdot \GL = 0$.

Notice that there is some arbitrariness in~$\GH$:
nothing changes if we add to it a linear 
combination of the primary Hamiltonian constraints because 
$\FL^*(\phi_\mu) = 0$ identically. 

In this paper we will consider the case where
the transformation itself is projectable to phase space, 
that is,
\beq
\deltaL q = \FL^*(\deltaH q)
\label{deltal} ,
\eeq
for a certain $\deltaH q(t,q,p)$.
Notice that there is also an arbitrariness in the determination
of $\deltaH q$
because of the existence of Hamiltonian constraints. 

Using $\GH$ and $\deltaH$, 
the Noether condition may be written
$$
[L]_i \,\FL^*(\deltaH q^i) + \deriv{\FL^*\GH}{t} = 0 ,
$$
from which, by extracting the coefficient of~$\ddot q$,
one obtains
$\ds
W \FL^*\left( \deltaH q -  \derpar{\GH}{p} \right) = 0
$.
From this equation, and using the null vectors of the Hessian,
it is easy to redefine $\GH$ and $\deltaH q$ conveniently
---using the primary Hamiltonian constraints---
in order to obtain 
\cite{BGGP89-noether}
\beq
\deltaH q^i = \frac{\partial \GH}{\partial p_i} = \{ q^i, \GH \} .
\label{deltah}
\eeq
In other words:
{\it a projectable Noether transformation  
is canonically generated in phase space}. 
On this basis we are ready to generalize equation (\ref{purecomm})
to the case of projectable Noether symmetries
associated with singular Lagrangian dynamics.
First we will give a characterization in phase space,
next we will give a characterization using the operator~$\K$,
and finally we will give a characterization in velocity space.

\subsection{Characterization in phase space}

Now we wish to study the Noether transformations in phase space. 
The dynamics of gauge theories, 
as examples of constrained systems in the Dirac sense, 
exhibit a certain amount of arbitrariness in order to account for
the gauge ---unphysical--- degrees of freedom. 
A typical evolution operator in phase space will be
\beq
\XH \approx  
{\partial \over \partial t} + \{ -, H \} + \lambda^\mu  \{ -, \phi_\mu \}
, 
\label{xh}
\eeq
where $\approx$ (Dirac's weak equality) 
is here an equality up to primary Hamiltonian constraints,
and $\lambda^\mu$ are a set of arbitrary Lagrange multipliers. 
As a matter of fact, 
these Lagrange multipliers are determined as functions in tangent space 
just by applying (\ref{xh}) to the configuration variables,
yielding $\lambda^\mu = v^\mu(q,\dot q)$
---see (\ref{K-H}).
 
Notice that the weak equality in (\ref{xh}) makes the definition of 
$\XH$ consistent with any redefinition of the basis of primary 
constraints.
However, this is not the final form of the dynamics.
To get the final dynamics we must perform a stabilization algorithm
\cite{Dir64,BGPR86-equiv,GNH78-pres,BK86-pres,Car90-theory}:
consistency requirements 
---that is, the tangency of $\XH$ to the surface of constraints--- 
may lead to new constraints and also
to the determination of some of the Lagrangian multipliers 
as functions in phase space.

Notice that, for any values we can give to the Lagrangian multipliers, 
the last piece in (\ref{xh}) may be written as $\{-, \phc\}$, 
where $\phc$ stands for an 
arbitrary linear combination of the primary Hamiltonian constraints,
\beq
\XH \approx  {\partial \over \partial t} + \{ -,  H \} + \{-, \phc\}.
\label{xh'}
\eeq

Let us consider the infinitesimal transformation generated by a
vector field $\VH$ in $\Tan^*Q$,
that is to say, $\deltaH h = \VH \cdot h$
---an infinitesimal parameter may be understood here.
The condition that $\VH$ be a symmetry of the dynamics 
is no longer characterized
by the strong condition of commutativity $[\VH, \XH]=0$. 
We may venture that the appropriate characterization is that 
the infinitesimal variation of $\XH$ produced by $\VH$,  
$$
\delta \XH = 
{\cal L}_{\VH} \XH =
[\VH, \XH] ,
$$ 
is of the type  $\{ -, \phc \}$, 
in order that the transformed vector field
is again of the type (\ref{xh'}). So, the characterization will read
\beq
[\VH , \XH] \approx  \{ -, \phc \} .
\label{qcomm}
\eeq

Since equation (\ref{xh'}) does not express the final form of the dynamics, 
we could produce more refined versions of (\ref{qcomm}). 
But, in the case of a Noether transformation, 
the invariance of the action is required not only on-shell but 
also off-shell, 
therefore the dynamics as given by 
(\ref{xh'}) is the right one to be used. 

Now let us prove that, when $\VH$ generates a canonical transformation, 
relation (\ref{qcomm}) is exactly 
the characterization of a projectable Noether transformation. 
We can write $\VH$ as
\beq
\VH = \{ -, \GH \}
\label{VG}
\eeq
for some function $\GH$,
so that $\deltaH h = \{h,\GH\}$. 
To eliminate the weak equalities in (\ref{xh'}),
$\XH$ can be written as
$$
\XH =  {\partial \over \partial t} + \{ -,  H \} + \{-, \phc\}
+ \phi_{\mu} \Z^{\mu}
$$
for some arbitrary vector fields $\Z^{\mu}$. 
Then, taking into account that
$$
[\XH, \VH] \approx \{-, \XH(\GH)+ \phc \} 
- \VH(\phi_{\mu}) \Z^{\mu},
$$
the requirement (\ref{qcomm}) becomes
$$
\VH(\phi_{\mu}) = \phc , \quad
\XH(\GH) = \phc + f(t),
$$
where $f(t)$ is an unknown function of time.
Notice that $\GH$ can be redefined by
$\ds \GH \to \GH - \int \!\!f(t) dt$,
since this does not change equation (\ref{VG}),
and hence we have
\beq
\VH(\phi_{\mu}) = \phc, \quad
\XH(\GH) = \phc ;
\label{tang}
\eeq
but since the functions $\lambda^\mu$ in 
the definition of $\XH$ (\ref{xh}) are arbitrary, 
the second equation in (\ref{tang}) splits into
\beq
{\partial \GH \over\partial t} + \{\GH,H\} =  \phc,
\label{cond2}
\eeq
and
\beq
\{\GH,\phi_{\mu} \} =  \phc.
\label{cond1}
\eeq
Notice that (\ref{cond1}) is just the first equation in (\ref{tang}).
 
It was proven in 
\cite{BGGP89-noether} that given a Noether transformation there exists
a function $\GH$ ---whose pullback to velocity space is the standard
conserved quantity $\GL$--- 
satisfying these conditions (\ref{cond2}) and (\ref{cond1});
and conversely, that these conditions ensure that the transformation
generated by $\GH$ through (\ref{deltah}) and (\ref{deltal})
is a Noether symmetry. 
What we have then obtained is a reformulation of 
(\ref{cond2}) and (\ref{cond1}) as commutativity conditions. 
To be more specific, we have proved the following result:

\begin{teor}
An infinitesimal transformation in phase space is 
a canonical Noether transformation 
if and only if its vector field $\VH$ satisfies
\beq
[\VH, \XH] \approx \{ -, \phc \}, 
\qquad
{\cal L}_{\VH} \Omega = 0 ,
\label{qcomm2}
\eeq
where $\XH$ is defined by (\ref{xh'}) and $\Omega$ is the 
symplectic form in phase space.
\end{teor}

(The contents of the second condition in (\ref{qcomm2}) is that 
$\VH$ generates canonical transformations.)

\subsection{Characterization using the evolution operator $\K$}

Now we will show an alternative characterization of Noether transformations
in phase space that makes use of a special evolution operator 
that connects the phase space picture with the velocity space picture. 
Gauge systems derived from a variational principle 
exhibit evolution vector fields, either in the Lagrangian formulation 
or in the Hamiltonian one, that contain some arbitrariness 
---because of the gauge freedom. 
But one can also consider a third evolution operator that, 
unlike the previous ones, is fully deterministic
\cite{BGPR86-equiv}. 
This is the operator $\K$ of section~2.

Using the operator $\K$, the Noether conditions 
(\ref{cond2}) and (\ref{cond1})
get the simpler form 
\cite{BGGP89-noether}
\beq
\K \cdot \GH = 0 .
\label{KG}
\eeq
Our scope is to present these Noether conditions in a new form, 
combining Hamiltonian and Lagrangian transformations
and involving commutations with both the pull-back operation and the 
evolution operator $\K$. 
This method has the advantage of its simplicity
because, as we said,
the operator $\K$ has none of the arbitrariness that plague
the evolution vector fields in velocity space and phase space.
In this sense, the commutation properties involving $\K$
will be the easiest ones to be used as a test of Noether symmetry.
In order to do so, we will prepare some preliminary results.

\medskip
First let us consider two infinitesimal transformations 
(leaving time invariant),
$\deltaH$ in phase space, and $\deltaL$ in velocity space.
In principle, they are unrelated, and do not necessarily describe symmetries.
For a function $h(t,q,p)$ the variation is computed 
in terms of $\deltaH q$ and $\deltaH p$ as
$$
\deltaH h(t,q,p) = \derpar hq \deltaH q + \derpar hp \deltaH p ,
$$
and similarly for a function $f(t,q,\dot q)$:
$$
\deltaL f(t,q,\dot q) = \derpar fq \,\deltaL q + \derpar f{\dot q} \,\deltaL \dot q .
$$ 
Using these relations, the definitions of $\FL$ and $\K$, and the chain rule,
a straightforward computation shows that
\bea
\deltaL \FL^*(h) - \FL^*(\deltaH h) &=&
\widehat{\derpar hq} (\deltaL q - \widehat{\deltaH q}) +
\widehat{\derpar hp} (\deltaL \widehat p - \widehat{\deltaH p}) ,
\label{comm-FL}
\\
\deltaL (\K \cdot h) - \K \cdot \deltaH h &=&
\left(\K \cdot \derpar hq\right) (\deltaL q - \widehat{\deltaH q}) +
\left(\K \cdot \derpar hp\right) (\deltaL \widehat p - \widehat{\deltaH p}) +
\nonumber
\\
&&
\widehat{\derpar hq} (\deltaL \dot q - \K \cdot \deltaH q) +
\widehat{\derpar hp} (\deltaL (\K \cdot p) - \K \cdot \deltaH p) ,
\label{comm-K}
\eea
where we have written $\widehat h$ for $\FL^*(h)$ to simplify the notation.
As a consequence, we have:

\begin{teor}
A necessary and sufficient condition in order that
$$
\deltaL (\K \cdot h) - \K \cdot \deltaH h = 0
$$ 
for each function~$h$,
is that the transformations $\deltaL$, $\deltaH$ be related by
\beasn
\deltaL q &=& \widehat{\deltaH q} \\
\deltaL \dot q &=& \K \cdot \deltaH q \\
\deltaL \widehat p &=& \widehat{\deltaH p} \\
\deltaL (\K \cdot p) &=& \K \cdot \deltaH p .
\label{relat}
\eeasn
Moreover, then one also has
$\deltaL \FL^*(h) - \FL^*(\deltaH h) = 0$.
\end{teor}

To prove the first assertion, one only has to take appropriate values for~$h$:
taking $h = q^i$ or $h = p_i$ leads to the vanishing 
of the last two terms in (\ref{comm-K});
taking $h = (q^i)^2/2$ leads to the vanishing of the first term;
finally, taking $h = q^i p_i$ (not summed) does the rest.

In view of this, the last assertion is a direct consequence
of (\ref{comm-FL}).
\medskip

From now on we suppose that 
the infinitesimal transformation in phase space is canonical,
and let $\GH(t,q,p)$ a generating function for it
(determined up to a function of time):
\beq
\deltaH q = \{q,\GH\} = \derpar \GH p, \quad 
\deltaH p = \{p,\GH\} = -\derpar \GH q .
\label{deltaH}
\eeq

We will need to know the partial derivatives of $\K \cdot h$.
A direct calculation from the definition (\ref{K}) yields
\bea
\derpar{(\K \cdot h)}{q} &=& 
\K \cdot \derpar hq + 
\dderpar Lqq \widehat{\derpar hp} +
\dderpar Lq{\dot q} \left( \K \cdot \derpar hp \right) ,
\label{Kh/q}
\\
\derpar{(\K \cdot h)}{\dot q} &=&
\widehat{\derpar hq} + 
\dderpar L{\dot q}q \widehat{\derpar hp} +
\dderpar L{\dot q}{\dot q} \left( \K \cdot \derpar hp \right) .
\label{Kh/v}
\eea
These relations
applied to $h = \GH$ yield
\bea
\derpar{(\K \cdot \GH)}{q} &=& 
-\K \cdot {\deltaH p} + 
\dderpar Lqq        \, \widehat{\deltaH q} +
\dderpar Lq{\dot q} \, (\K \cdot \deltaH q) ,
\label{KG/q}
\\
\derpar{(\K \cdot \GH)}{\dot q} &=&
-\widehat{\deltaH p} + 
\dderpar L{\dot q}q        \, \widehat{\deltaH q} +
\dderpar L{\dot q}{\dot q} \, (\K \cdot \deltaH q) .
\label{KG/v}
\eea

Now let us write $\deltaL f$
for $f = \K \cdot p = \derpar Lq$
and for $f = \widehat p = \derpar L{\dot q}$.
We obtain the identities
\beann
0&=& 
\deltaL (\K \cdot p) - 
\dderpar L{q}{q}      \, \deltaL q - 
\dderpar L{q}{\dot q} \, \deltaL{\dot q}
,
\\
0&=& 
\deltaL \widehat p - 
\dderpar L{\dot q}{q}      \, \deltaL q - 
\dderpar L{\dot q}{\dot q} \, \deltaL{\dot q}
.
\eeann
Using these relations,
equations (\ref{KG/q}) and (\ref{KG/v}) become
\bea
\derpar{(\K \cdot \GH)}{q} &=& 
\deltaL (\K \cdot p) - \K \cdot {\deltaH p} + 
\dderpar Lqq        (\widehat{\deltaH q} - \deltaL q) +
\dderpar Lq{\dot q} (\K \cdot \deltaH q - \deltaL{\dot q}) ,
\label{KG/q'}
\\
\derpar{(\K \cdot \GH)}{\dot q} &=&
\deltaL \widehat p - \widehat{\deltaH p} + 
\dderpar L{\dot q}q        (\widehat{\deltaH q} - \deltaL q) +
\dderpar L{\dot q}{\dot q} (\K \cdot \deltaH q - \deltaL{\dot q}) .
\label{KG/v'}
\eea

So far we have not made any assumption on the relationship between $\deltaH$ and~$\deltaL$,
but from the preceding equations the following result is clear:

\begin{teor}
Let $\GH(t,q,p)$ be the generator of an infinitesimal transformation
$\deltaH$ in phase space (\ref{deltaH}).
If we define an infinitesimal transformation $\deltaL$ in velocity space by
\beq
\deltaL q := \FL^*(\deltaH q) ,
\qquad
\deltaL {\dot q} := \K \cdot \deltaH q
,
\label{deltaL-H}
\eeq
then we have
\bea
\derpar{(\K \cdot \GH)}{q} &=& 
\deltaL \K \cdot p - \K \cdot {\deltaH p} ,
\label{KG/q''}
\\
\derpar{(\K \cdot \GH)}{\dot q} &=&
\deltaL \FL^*(p) - \FL^*(\deltaH p) .
\label{KG/v''}
\eea
\end{teor}

Under the assumptions of the theorem,
we can rewrite the commutation relations
(\ref{comm-FL}), (\ref{comm-K}) as
\bea
\deltaL \FL^*(h) - \FL^*(\deltaH h) &=&
\widehat{\derpar hp} \derpar{}{\dot q}(\K \cdot \GH) ,
\label{comm-FL'}
\\
\deltaL (\K \cdot h) - \K \cdot \deltaH h &=&
\left( \K \cdot \derpar hp \right) \derpar{}{\dot q}(\K \cdot \GH) +
\widehat{\derpar hp} \derpar{}{q}(\K \cdot \GH) .
\label{comm-K'}
\eea

The final step is to relate these relations with the condition (\ref{KG}), 
$\K \cdot \GH = 0$,
that characterizes the generators of projectable Noether transformations:

\begin{teor}
Let $\deltaH$ be a canonical transformation in phase space,
and let $\deltaL$ be defined as in
(\ref{deltaL-H}).
Then the following statements are equivalent:
\vadjust{\kern-2mm}
\ben
\itemsep 0pt
\parskip 0pt plus 1pt
\item
The commutation relation
$\deltaL (\K \cdot h) - \K \cdot \deltaH h = 0$ 
holds for each function~$h(t,q,p)$.
\item
$\deltaH$ is a Noether transformation in phase space.
\een
\end{teor}

To prove that the first condition implies the second one,
notice that, using theorems 2 and~3,
if $\GH$ is a generator of~$\deltaH$,
$\K \cdot \GH$ is a function of time, $f(t)$.
Redefinition of $\GH$ to $\ds \GH - \int \!\!f(t)$
makes $\K \cdot \GH = 0$,
therefore, according to (\ref{KG}), 
$\deltaH$ is a Noether transformation in phase space.
The converse is a direct consequence of (\ref{KG}) and theorem~3.

\medskip

Let us finally remark that we could have defined, 
instead of (\ref{deltaL-H}), 
\beq
\bar\deltaL q := \FL^*(\deltaH q) ,
\qquad
\bar\deltaL {\dot q} := 
\deriv{}{t} \bar\deltaL q . 
\label{deltaL-H'}
\eeq
Here the Lagrangian transformation of $q$ is the pull-back of the Hamiltonian one,
whereas the transformation of the velocity 
is the natural prolongation of the transformation of the position.
This is the usual way to define the transformations of the velocities
out of the transformations of the positions.
Notice that, using equation (\ref{K-EL}),
$$
\bar\deltaL \dot q^i =
\deltaL \dot q^i - [L]_j \,\FL^*\left( \derpar{\deltaH q^i}{p_j} \right) ,
$$
so both transformations coincide when applied to 
solutions of the Euler-Lagrange equation
(they coincide ``on-shell'').
With $\bar\deltaL$ instead of $\deltaL$ equations 
(\ref{KG/q''}) and (\ref{KG/v''}) acquire an additional term 
that vanishes on-shell.
Therefore $\deltaL$ as defined in theorem~3
is more appropriate in order to give a neat characterization 
of a Noether transformation through commutation relations. 
Nevertheless, it is $\bar\deltaL$ that, when applied to the
Lagrangian, gives a total derivative. Indeed, from (\ref{noether}),
one has
$$
\bar\deltaL L = \deriv{}{t}\FL^*( p \; \deltaH q - \GH).
$$ 



\subsection{Characterization in velocity space}

To obtain a characterization in velocity space 
we first need to formulate the dynamics as 
a vector field in $\Real \times \Tan Q$. 
The time evolution in a gauge theory is not unique until the gauge
freedom has been removed 
---by way of some gauge fixing, for example. 
This is reflected in the ambiguities that are present
in the Lagrangian time-evolution differential operator, 
which we recall from section~2:
$$
\XL = 
\X_0 + \eta^{\mu} \Gamma_{\mu} ,\quad
\X_0 =
\derpar{}{t}
+ \dot q^{i} {\partial\over\partial q^{i}}
+ a^{i}(q,\dot q) {\partial\over\partial \dot q^{i}} .
$$
Notice that projectable quantities have, according to (\ref{projectable}), 
a well-defined unambiguous time-derivative under this dynamics.
The requirement of tangency of $\XL$ to the primary Lagrangian 
constraint submanifold, defined by 
$\chi_\mu \approx 0$
(\ref{primlag}),
may lead to new constraints and to the determination of some of the 
functions $\eta^\mu$. 
At this point, new tangency requirements may occur. 
This is the Dirac's method in the Lagrangian formalism
\cite{BGPR86-equiv}.

\smallskip
Our aim is to give a tangent space characterization of a Noether 
transformation $\deltaL q(t; q, \dot q)$ that satisfies the property of
being projectable to phase space, that is, 
$\deltaL q$ is the pullback of a canonical Noether transformation 
$\deltaH q$, 
$\deltaL q  = \FL^*(\deltaH q)$. 
Notice at this point that we have two natural ways to define 
the dynamical time derivative $\deltaL \dot q$ in 
$\Real \times \Tan Q$. 
Either by 
$\deltaL \dot q := \K \cdot \deltaH q$ 
as in the preceding subsection, or by 
$\deltaL \dot q := \XL \cdot \deltaL q = \X_0 \cdot \deltaL q$. 
According to (\ref{XandK}), both definitions coincide only on the primary Lagrangian constraints submanifold.
Consistency with the preceding subsection invites us to choose the definition 
$\deltaL \dot q := \K \cdot \deltaH q$, 
and this is what we will do. 
So we take
\beq
\VL = \FL^*\{q,\GH\} \,\derpar{}{q} + \K \cdot \{q,\GH\} \,\derpar{}{\dot q} .
\label{vl}
\eeq 

We will use the results of the preceding subsection for Noether 
transformations, in particular
\beq
\K  \cdot \deltaH h = \deltaL (\K \cdot h) 
\label{kcom}
\eeq
and its consequence
\beq
\FL^* ( \deltaH h) = \deltaL \FL^*(h), 
\label{flcom}
\eeq
for any function $h$ on $\Real \times \Tan^*Q$.

Notice from these relations that
$$
\VL \cdot \FL^*(h) = \FL^*\{h,\GH\} :
$$
the action of $\VL$ on a projectable function
is a projectable function,
that is, 
$\VL$ is a projectable vector field
---indeed it projects to $\VH = \{-,\GH\}$.

\smallskip
Equation (\ref{kcom}), and the fact that the primary Lagrangian constraints 
($\plc$) can be obtained as 
$\chi_\mu = \K \cdot \phi_\mu$,
allow to compute
$$
\VL \cdot \chi_\mu =
\deltaL \chi_\mu = \deltaL (\K \cdot \phi_\mu) = 
\K \cdot (\deltaH \phi_\mu) = \K \cdot \{ \phi_\mu,  \GH\},
$$
but, according to (\ref{cond1}), 
\beq
\{ \phi_\mu,  \GH\} = D^\nu_\mu \,\phi_\nu
\label{cond1'}
\eeq
for some functions $D^\nu_\mu$. 
Therefore 
$$
\VL \cdot \chi_\mu =
\deltaL \chi_\mu  = \FL^*(D^\nu_\mu) \,\chi_\nu ,
$$
that is: 
$\VL$ is tangent to the primary Lagrangian constraints surface~$V_1$,
$\VL (\plc) = \plc$.

Now let us use (\ref{XandK}) and (\ref{kcom}) to write
$$
\VL (\X_0 \cdot \FL^*(h)) + 
\VL \left(
  \chi_\mu \,\derpar{v^\mu}{\dot q} \FL^*\left(\derpar{h}{p}\right)
  \right) 
= 
\X_0 \cdot \FL^*(\deltaH h) +
  \chi_\mu \derpar{v^\mu}{\dot q} \FL^*\left(\derpar{\deltaH h}{p}\right) .
$$
The second piece in the right side is just a combination of $\plc$, 
and so it is the second piece in the left side 
because of the tangency of $\VL$ to the $\plc$ surface. 
Therefore
$$
\VL (\X_0 \cdot \FL^*(h)) - \X_0 \cdot \FL^*(\deltaH h) = \plc ,
$$
or, using (\ref{flcom}), 
\beq
[\VL, \X_0] (\FL^* h) = \plc .
\label{dellxcom0}
\eeq

This result means that the commutator $[\VL, \X_0]$ is,
on the $\plc$ surface, 
a combination of the vector fields in the kernel of $\Tan(\FL)$, 
that is,
\beq
[\VL, \X_0] = \plc + \alpha^\mu \Gamma_\mu ,
\label{dellxcom}
\eeq
for some functions $\alpha^\mu$.

We need a second piece of information: 
the commutator 
$[\VL, \Gamma_\mu]$. 
Let us apply it to a configuration variable~$q$. 
Since 
$\Gamma_\mu \cdot q = 0$ and 
$\Gamma_\mu \cdot \deltaL q = \Gamma_\mu \cdot \FL^*(\deltaH q) = 0$, 
we get 
$[\VL, \Gamma_\mu] \cdot q = 0$. 
When applied to $\dot q$,  
$$
[\VL, \Gamma_\mu] \cdot \dot q = 
\VL \gamma_\mu - \Gamma_\mu \VL(\dot q) = 
\VL \gamma_\mu - \Gamma_\mu \cdot (\K \cdot \deltaH q) ,
$$
where in the last step we have used the definition 
$\VL \cdot \dot q = \K \cdot \deltaH q$. 
Taking into account the definition (\ref{gamma}) and the property 
(\ref{Gamma-K}),
we get
\bea
[\VL, \Gamma_\mu] \cdot \dot q
&=& 
\VL(\FL^*\{q,\phi_\mu\}) - \FL^*\{\deltaH q,\phi_\mu\} 
\\
&=& 
\FL^*(\deltaH\{q,\phi_\mu \} - \{\deltaH q,\phi_\mu\})
 = \FL^*\{q,\deltaH \phi_\mu\}.
\eea
We can use again (\ref{cond1'}), 
$\deltaH \phi_\mu = \{\phi_\mu, \GH \} = D^\nu_\mu \,\phi_\nu$. 
Then,
$$
[\VL, \Gamma_\mu] \cdot \dot q
= \FL^*\{ q, D^\nu_\mu \phi_\mu \}
= (\FL^* D^\nu_\mu) \, \gamma_\nu,
$$
and therefore
\beq
[\VL, \Gamma_\mu] 
= (\FL^* D^\nu_\mu) \Gamma_\nu ,
\label{delgamcom}
\eeq
which agrees with the fact that $\VL$ is projectable.

Putting together (\ref{dellxcom}) and (\ref{delgamcom}), 
we obtain that the vector field $\VL$ satisfies
\beq
[\VL, \XL] = \plc + \beta^\mu \Gamma_\mu 
\label{vl-xl}
\eeq
for some functions $\beta^\mu(t;q,\dot q)$.

So we have proved the following result:

\begin{teor}
Suppose that $\GH(t,q,p)$ generates a canonical Noether transformation,
and let $\VL$ be the vector field defined by it 
according to (\ref{vl}).
Then $\VL$ is a projectable vector field
that projects to $\{-,\GH\}$,
it is tangent to the primary Lagrangian constraint submanifold,
and its commutation with the dynamics satisfies
(\ref{vl-xl}).
\end{teor}

This result is analogous to that of subsection 3.1. 
Here and there the commutator of the generator of the
transformation with the evolution vector field 
gives as a result a term which is proportional 
to the arbitrary piece in the dynamics. 
Have we reached a necessary
and sufficient condition for $\VL$ to be a generator of a 
projectable Noether transformation? The answer in general is in the negative.
Let us be more specific and consider a vector field $\VL$, 
defined in (\ref{vl}), such that: 
(a) it projects to $\{-,\GH\}$, 
(b) is tangent to the primary lagrangian constraint submanifold, and 
(c) satisfies (\ref{vl-xl}). 
Then, using equations (\ref{comm-FL'}) and (\ref{comm-K'}) one arrives at
\beq
\derpar{}{\dot q}(\K \cdot \GH) = 0, \qquad
\derpar{}{q}(\K \cdot \GH) = \plc,
\label{nonecsuf} 
\eeq
whereas the right conditions for $\{-,\GH\}$ to generate a Noether 
transformation in phase space 
---which implies that $\VL$ generates a Noether 
transformation in tangent space--- 
are, according to the discussion in the preceding section, 
\beq
\derpar{}{\dot q}(\K \cdot \GH) = 0, \qquad
\derpar{}{q}(\K \cdot \GH) = 0,
\label{necsuf}
\eeq
which is more restrictive than (\ref{nonecsuf}). However, in most cases of
interest, the $\plc$ do not restrict the configuration
variables alone, and then (\ref{nonecsuf}) and (\ref{necsuf}) are equivalent. 
This
is the case indeed in many physical applications of gauge systems, as in 
string theory, Yang--Mills theory, or general relativity. 
In such cases we have arrived at a characterization of
the vector field $\VL$ for it to generate a Noether transformation. 

The case where the $\plc$ do restrict the configuration variables 
is rather unusual, and it might be considered as an unfortunate choice
of the configuration space
---some comments on this issue can be found in
\cite{MMS-constraints}.
The second example in the following section, 
though formal and with no physical interest,
exhibits this feature;
in this case, 
conditions (\ref{nonecsuf}) are not sufficient
for $\VL$ to generate a Noether transformation.

Let us finally recall that the action of the vector field $\VL$, associated
to a projectable Noether transformation, on 
the Lagrangian $L$ does not give in general a total derivative. 
The transformation that indeed gives a total derivative
is $\bar \deltaL$
---see the end of subsection 3.2.

\subsection{The algebra of projectable Noether symmetries}

Consider a canonical Noether symmetry generated by $\GH$.
The projectability of (\ref{vl}),
$$
\VL = \FL^*\{q,\GH\} \,\derpar{}{q} + \K \cdot \{q,\GH\} \,\derpar{}{\dot q}, 
$$
to the canonical generator of Noether symmetries,
$$
\VH = \{ -,\GH\},
$$
allows to obtain some results concerning the algebra of the vector fields 
associated to projectable Noether symmetries.
If $\VL_{1}$ and $\VL_{2}$ are two such vector fields, 
associated with the canonical generating functions $\GH_{1}$ and $\GH_{2}$, 
then it is straightforward to show that the commutator
$
[\VL_{1}, \VL_{2}]
$
projects to 
$
\{-, \{\GH_{2},\GH_{1}\} \}
$
that is,
$$
[\VL_{1}, \VL_{2}] \cdot \FL^*(h) = \{h, \{\GH_{2},\GH_{1}\} \} .
$$

In the particular case that the set of independent canonical 
generators $G_{i}$ span a Lie algebra,
\beq
\{\GH_{i},\,\GH_{j} \} = C_{ij}^{k}\GH_{k},
\label{algebr}
\eeq
with $C_{ij}^{k}$ constants,
then their associated vector fields $\VL_{i}$ in tangent space satisfy 
the same Lie algebra structure,
$$
[\VL_{i},\VL_{j}] = C_{ij}^{k} \VL_{k}.
$$

In the case that the quantities in (\ref{algebr}) are not constants 
but functions of the variables 
(this is the case of a ``soft'' algebra generating a ``quasigroup'' 
\cite{Bat81}), this last equality does not hold, 
but we still have the opportunity to get 
---up to pieces linear in the primary constraints--- 
the structure functions in phase space by Lagrangian methods. 
This goes as follows.
Consider (\ref{algebr}) for some functions $C_{ij}^{k}$. 
Consider also the pullback to tangent space of the 
canonical generating functions, $\GL_{i} = \FL^*(\GH_{i})$. 
Then
\bea
\VL_{j} \cdot \GL_{i} &=& 
 \VL_{j} \cdot \FL^*(\GH_{i}) = 
 \FL^* (\deltaH_{j} \GH_{i}) =
 \FL^* \{\GH_{i},\GH_{j}\} 
\nonumber 
\\
&=& 
\FL^*(C_{ij}^{k} \GH_{k}) = \FL^*(C_{ij}^{k}) \GL_{k} .
\eea
That is, we can retrieve ---up to primary constraints--- 
the structure functions of the canonical gauge generators 
by simply computing the variations under the vector fields $\VL$ 
of the Noether conserved quantities in tangent space. 
This method has been implicitly used in a series of papers 
\cite{larrydon1,larrydon2,larrydon3} 
that analyze the relationship between 
the Lagrangian and Hamiltonian descriptions 
of the gauge group structure for generally covariant theories.

\section{Some examples}

\paragraph{Example 1.}
Let us consider the Lagrangian
$$
L = \frac12 e^{-\omega} \dot\bfx^2 + \frac12 e^\omega m^2 ,
$$
which describes a free particle in Minkowski's space-time.
A standard analysis yields the momenta 
$(\bfp,\pi)$ of the variables $(\bfx,\omega)$,
a Hamiltonian function, and a primary Hamiltonian constraint:
$$
\widehat \bfp = e^{-\omega} \dot\bfx , \quad 
\widehat \pi = 0 , \quad
H = \frac12 e^\omega (\bfp^2-m^2) , \quad
\phi^0 = \pi .
$$
The stabilization algorithm yields a secondary Hamiltonian constraint
$$
\phi^1 = \{\phi^0,H\} = -H .
$$
The evolution operator $\K$ is given by
$$
\K \cdot h  = 
\FL^*\derpar{h}{t} + \dot\bfx \, \FL^*\derpar{h}{\bfx} 
+ \dot\omega \,\FL^*\derpar{h}{\omega}
+ \chi \,\FL^*\derpar{h}{\pi} ,
$$
where we have denoted by $\chi$ the primary Lagrangian constraint
$$
\chi := \K \cdot \phi^0 = \frac12 (e^\omega m^2 - e^{-\omega} \dot\bfx^2) .
$$
It is clear that the projectable functions are those not depending
on $\dot\omega$, and indeed the kernel of $\Tan(\FL)$ is spanned by
$$
\Gamma = \derpar{}{\dot\omega} .
$$
Notice therefore that $\chi$ is a projectable constraint, and so 
$$
\chi = \FL^*(\phi^1) ,
$$
whereas
$$
\K \cdot \phi^1 = \dot\omega \chi ,
$$
which is not a new constraint.
Finally, we give the Euler-Lagrange equations:
$$
[L]_\bfx = e^{-\omega}(\dot\omega \dot\bfx - \ddot\bfx) ,
\quad
[L]_\omega = \chi .
$$

At the first stage of the stabilization algorithm
the Hamiltonian evolution operator is
$$
\XH = \derpar{}{t} + e^\omega \bfp \derpar{}{\bfx} - \frac12 e^\omega (\bfp^2 - m^2) \derpar{}{\pi} + \lambda \derpar{}{\omega} + \pi \Z ,
$$
where the function $\lambda$ and the vector field $\Z$ are arbitrary.

Now let us study the gauge transformations. 
From the general theory, a gauge generator has the form 
$\GH = \dot\eps G_0 + \eps G_1$,
where $\eps$ is an arbitrary function of time
and the functions $G_i$ are determined such that $\K \cdot \GH = 0$
\cite{GP88-ggen}.
$G_0$ is a first-class primary Hamiltonian constraint,
which in this example turns out to be
$e^{-\omega} \pi$.
The result is
$$
\GH = e^{-\omega} (\dot\eps \phi^0 - \eps \phi^1) .
$$
Its associated infinitesimal transformation is given by the vector field
$$
\VH = \eps \bfp \derpar{}{\bfx} + \dot\eps e^{-\omega} \derpar{}{\omega} +
\dot\eps e^{-\omega} \pi \derpar{}{\pi} .
$$
Let us check the quasi-invariance of $\XH$:
$$
[ \VH,\XH] = 
\left( e^{-\omega}(-\ddot\eps + \lambda \dot\eps) +  \VH \cdot \lambda \right)
\derpar{}{\omega} +
\pi \left( e^{-\omega}(-\ddot\eps + \lambda \dot\eps)\derpar{}{\pi} 
+ [ \VH, \Z ] + e^{-\omega} \dot\eps \Z \right) ,
$$
which is weakly $\{-,\phc\}$.

The vector field $\VL$ of (\ref{vl}) is
$$
 \VL = e^{-\omega} \left(
 \eps \bfx \derpar{}{\bfx} +
 \dot\eps \derpar{}{\omega} + 
 \dot\eps \dot\bfx \derpar{}{\dot\bfx} +
 (\dot\eps \dot\omega - \ddot\eps e^{-\omega}) \derpar{}{\dot\omega} 
\right) .
$$
A direct computation then shows that, as differential operators,
$$
 \VL \circ \K - \K \circ  \VH = 0 .
$$

Finally let us consider the Lagrangian dynamical vector field,
$$
\XL = \derpar{}{t} + 
\dot\bfx \derpar{}{\bfx} + \dot\omega \derpar{}{\omega} +
\dot\omega \dot\bfx \derpar{}{\dot\bfx} + \eta \derpar{}{\dot\omega} ,
$$
where $\eta$ is an arbitrary function.
Then we obtain
$$
[ \VL,\XL] =
e^{-\omega} \left( -\dddot\eps + 2\ddot\eps\dot\omega -
 \dot\eps \dot\omega^2 + \dot\eps e^{-\omega} \eta +  \VL \cdot \eta \right)
\derpar{}{\dot\omega} ,
$$
which is proportional to~$\Gamma$.

Moreover, bearing in mind the remarks at the end of section 3.2,
we can define the Lagrangian transformation of the velocities
as the time-derivatives of the transformation of the positions,
thus obtaining a slightly different vector field $\bar \VL$; 
indeed,
$$
\bar  \VL =  \VL - \eps [L]_{\bfx} \derpar{}{\dot\bfx} .
$$
Then the Noether condition can be checked for this transformation:
$$
\bar  \VL \cdot L = \deriv{}{t}( \eps e^{-\omega} L ) .
$$

\paragraph{Example 2.}
Here we show that, in general,
the conditions stated by theorem~5 are not sufficient
for $\VL$ to define a Noether transformation.
Let us consider 
$$
L = \frac12 \dot x^2 - \frac12 y^2 .
$$
The momenta 
$(p_x,p_y)$ of the variables $(x,y)$,
a Hamiltonian function and a primary Hamiltonian constraint are
$$
\widehat p_x = \dot x , \quad 
\widehat p_y = 0 , \quad 
H = \frac12 p_x^2 + \frac12 y^2 , \quad
\phi^0 = p_y .
$$
The stabilization algorithm yields a secondary Hamiltonian constraint
$$
\phi^1 = \{\phi^0,H\} = -y .
$$
The evolution operator $\K$ is given by
$$
\K \cdot h = \FL^*\derpar{h}{t} + \dot x \,\FL^*\derpar{h}{x} 
+ \dot y \,\FL^*\derpar{h}{y}
- y \,\FL^*\derpar{h}{p_y} .
$$
Notice that there are a primary Lagrangian constraint and a secondary one,
$$
\chi^1 = -y , \quad
\chi^2 = -\dot y .
$$

The projectable functions are those not depending
on $\dot y$, and the kernel of $\Tan(\FL)$ is spanned by
$$
\Gamma = \derpar{}{\dot y} .
$$
The Euler-Lagrange equations are
$$
[L]_x = - \ddot x ,
\quad
[L]_y = -y ,
$$
and so the Lagrangian evolution operator may be taken as
$$
\XL_0 = \dot x \derpar{}{x} + \dot y \derpar{}{y} .
$$

Let us consider the function
$$
\GH = p_y y ,
$$
whose associated infinitesimal transformation is the vector field in phase space
$$
\VH = y \derpar{}{y} - p_y \derpar{}{p_y} ,
$$
and defines the vector field in tangent space
$$
\VL = y \derpar{}{y} + \dot y \derpar{}{\dot y} .
$$
It is easily checked that
$\VL$ projects to $\VH$.
It is clear that $\VL \cdot \chi^1 = \chi^1$,
and so it is tangent to the primary Lagrangian constraint submanifold.
And also we have
$
[\VL,\XL_0] = 0 
$.

In spite of satisfying these three conditions of theorem~5,
$\VL$ is not a projectable Noether transformation.
We can see this in several ways.
On the one hand,
$\K \cdot \GH = -y^2$,
which is not zero
(notice, however, that since this is a primary lagrangian constraint
then $\GH$ corresponds to a nonprojectable Noether transformation,
see
\cite{GP92-hamnoeth} and
\cite{GP99-noeth}).
On the other hand, we can compute
$$
 (\VL \circ \K - \K \circ  \VH) \cdot h = -2y \widehat{\derpar{h}{p_y}} ,
$$
which is not zero.
Finally, using the transformation $\bar \VL$ as before, we have
$$
\bar  \VL \cdot L = -y^2 ,
$$
which is not a total derivative.

Finally,
we use this example to illustrate item (b) in the list of
properties of gauge theories given in the introduction.
Take the conserved quantity 
$\GH = p_x + p_y y$ in phase space.
It generates, through Poisson bracket,
an infinitesimal symmetry transformation $\deltaH$
whose pull-back to velocity space is 
$\deltaL x = 1$, $\deltaL y = y$;
this gives $\deltaL L = -y^2$,
which is not a total derivative
and thus $\deltaL$ is not a Noether symmetry.

\section{Conclusions}

In this paper we have introduced some characterizations of Noether symmetries based upon some specific properties of commutativity
with the dynamics. 
This presentation entails a shift of focus with respect to 
the standard introductions to Noether symmetries.

To our knowledge,
the only characterization of Noether symmetries in gauge theories,
not relying on properties of the conserved quantity,
is the invariance of the action under these transformations.
Our contribution is a new characterization of such symmetries
which is set up in the realm of dynamics, either Lagrangian or Hamiltonian.
This study concerns those Noether symmetries 
that are projectable to phase space 
(what we call canonical Noether transformations).

For canonical Noether symmetries we obtain a characterization
in phase space
that clearly generalizes the results that hold for regular
(not gauge) theories. 
We also provide an alternative characterization by using
the unambiguous evolution operator that connects the formulations 
in phase space and in tangent space;
this new characterization is very appropriate because of its simplicity, 
since it is set up with the only use of the Lagrangian function and
its partial derivatives. 
Finally, we give a characterization in velocity space applicable
to most dynamical theories with physical contents.  

In summary, 
we give an answer to the question of extending
the property of commutation of the Noether symmetry with the dynamics,
as expressed by equation~(\ref{purecomm}),
to singular Lagrangians. 
This answer is presented as three characterizations 
that may serve as a useful test of Noether symmetry for gauge theories
with reference neither to the action nor to the conserved quantity.

\subsection*{Acknowledgements}
X.\,G acknowledges financial support by CICYT project TAP 97-0969-C03. 
J.\,M.\,P acknowledges financial 
support by CICYT, AEN98-0431, and CIRIT, GC 1998SGR.



\end{document}